\documentclass[prb,twocolumn,showpacs,showkeys,amsmath,amssymb]{revtex4}
\usepackage{graphicx}% Include figure files
\usepackage{dcolumn}% Align table columns on decimal point
\usepackage{bm}% bold math
\newcommand{\rmmax}{\mathrm{max}}
\newcommand{\rmmin}{\mathrm{min}}

\newcommand{\dd}{{\mathrm{d}}}

\newcommand{\bfk}{{\mathbf{k}}}

\newcommand{\bfr}{{\mathbf{r}}}
\newcommand{\bfR}{{\mathbf{R}}}

\newcommand{\Hcal}{{\cal H}}

\begin{document}
\title{Boundary conditions for augmented plane wave methods}
\author{Christian Brouder}
% \email{antoine@lmcp.jussieu.fr}
% \homepage{http://www.lmcp.jussieu.fr/~antoine}
\affiliation{Insitut de min\'eralogie et de physique des milieux
    condens\'es,
    UMR CNRS 7590, Universit\'e Pierre et Marie Curie,
    case 115, 4 place Jussieu,
    F-75252 Paris Cedex 05, France}
\date{\today}
\begin{abstract}
The augmented plane wave method uses the Rayleigh-Ritz principle
for basis functions that are continuous but with discontinuous
derivatives and the kinetic energy is written as a pair of
gradients rather than as a Laplacian. It is shown here that
this procedure is fully justified from the mathematical point of view.
The domain of the self-adjoint Hamiltonian, which does not contain
functions with discontinuous derivatives, is extended to its
form domain, which contains them, and this modifies the form of
the kinetic energy. 
Moreover, it is argued that discontinuous basis functions 
should be avoided.
\end{abstract}
%71.15.Mb Density functional theory, local density approximation,
\maketitle
\section{Introduction}
The augmented plane wave (APW) method and its by-products are
among the most powerful electronic structure calculation 
schemes \cite{Singh}.
The augmented plane wave method was invented
by Slater in 1937 \cite{Slater37}.
Its basic idea is to partition the crystal
into two types of regions: (i) non-overlapping spheres $S_\alpha$
centered around all constituent atomic sites 
and (ii) the remaining interstitial region.
The solutions $\psi(\bfr)$ of the Schr\"odinger equation are
expanded over basis wavefunctions
$\phi_i(\bfr)$ which are plane waves in the interstitial
region and spherical waves \cite{ftn1}
inside each sphere.
The coefficients of this expansion are obtained through
the Rayleigh-Ritz principle \cite{Rayleigh,Ritz} by finding
the extrema of the Rayleigh quotient \cite{ftn2}
\begin{eqnarray}
E &=& \frac{\langle \psi | H | \psi\rangle}
                   {\langle \psi | \psi\rangle}.
\label{Rayleighquot}
\end{eqnarray}

A subtle point of the APW method is the boundary conditions for 
the basis functions $\phi_i$ at the surface of the spheres $S_\alpha$. 
Various choices can be found in the literature. In the linearized APW 
method \cite{Andersen}, the basis functions and their derivatives are
continuous at the spheres. In the original APW method, as
well as in the recent linearized APW method with local orbitals
\cite{Sjostedt,Madsen}, the basis functions are only continuous
at the spheres. Moreover, many authors proposed to use
basis functions that are discontinuous at the spheres
\cite{Brownstein,Szmytkowski}.

The boundary conditions are important to define the basis functions
but also to define the Rayleigh quotient itself. If the functions and
their derivatives are continuous, an integration by parts
(or Green's first identity)
shows that the kinetic energy term of the Hamiltonian can be
written in two equivalent forms:
\begin{eqnarray}
-\int\phi_i^*(\bfr)\Delta\phi_j(\bfr)\dd\bfr &=&
\int(\nabla\phi_i(\bfr))^*\cdot\nabla\phi_j(\bfr)\dd\bfr.
\label{lap=grad}
\end{eqnarray}
If the derivatives of the basis functions are not continuous,
then the left-hand side (i.e. the Laplacian form)
and the right-hand side (the gradient form) of 
equation (\ref{lap=grad}) are generally different and the question is:
which one (if any) is correct? Slater's answer
is ``the second, more fundamental form is the correct one, as it is the
one which directly enters the variation principle from 
which Schr\"odinger's equation is derived'' \cite{Slater37}.
Obviously, this is not a very convincing argument because the
variational principle can be written with the kinetic energy
in the Laplacian form as well as in the gradient form.
Still, this argument (and practical success) has been up to now
the only one to justify the use of the gradient form.
If the basis functions are not continuous, additional terms
are introduced \cite{Brownstein,Szmytkowski} but 
the method looses many of the attractive features of the
standard APW approach.

The purpose of this paper is to show that the question of the
boundary conditions in the APW method is not a matter of physical
principles, but a well defined mathematical question with
a precise answer which is:
basis functions with discontinuous derivatives are valid
if the gradient form is used and discontinuous basis functions
should be employed with much caution because they do not allow for a correct 
variational principle.

To obtain this answer, we make a short trip in the
mathematical theory of Hamiltonian operators.
We first consider the fact that realistic Hamiltonians $H$ are defined on 
a subspace $D(H)$ of the Hilbert space called the domain of $H$.
The Rayleigh-Ritz principle holds for basis functions in the domain of $H$.
However, the domain of $H$ does not contain functions with discontinuous
derivatives. To extend the possible basis functions, we use the
form domain $Q(H)$ of $H$. We show that the Rayleigh-Ritz principle
is still valid on $Q(H)$ if the gradient form of the kinetic energy
is used. These results fully confirm Slater's intuitions.
Finally, we show that the Rayleigh-Ritz principle is not valid if
the basis functions are not continuous. This implies a number of
unpleasant consequences which show that discontinuous basis
functions should be used with great care.

\section{The Hamiltonian and its domain}
It is sometimes thought that a Hamiltonian is a Hermitian operator 
$H$ defined on a Hilbert space $\Hcal$ in the sense that for any
$\phi$, $\psi$ in $\Hcal$ we have
$\langle H\phi|\psi\rangle= \langle \phi|H\psi\rangle$.
This not true for realistic Hamiltonians.
The Hellinger-Toeplitz theorem
(ref. \onlinecite{ReedSimonI} p.84) tells us that an operator
$H$ can satisfy $\langle H\phi|\psi\rangle= \langle \phi|H\psi\rangle$
for all $\phi$, $\psi$ in $\Hcal$ if and only if
the eigenenergies of $H$ are between 
a minimum and a maximum finite energy. Realistic Hamitonians have
a minimum energy (that of the ground state) but they
have no maximum energy. From the physical point of view, 
a system represented
by a Hamiltonian with a maximum energy would completely reflect
photons or electrons above that energy.  Experiment tells us
that very high energy photons and electrons go through matter
so such Hamiltonian is not realistic. From the mathematical
point of view, the Hamiltonians made of a kinetic energy term
plus a potential energy term that is periodic or go to zero at infinity
(plus some regularity assumptions) has no maximum energy. 

As a consequence, a realistic Hamiltonian $H$ is not defined on
a Hilbert space $\Hcal$ but on a subspace $D(H)$ of $\Hcal$, called
the domain of $H$, such that $H|\phi\rangle\in\Hcal$
for any $\phi\in D(H)$. We shall see that the question of the boundary
conditions turns out to be a question concerning the domain of $H$.

\subsection{Self-adjoint Hamiltonian}
Since the difference between the Laplacian and the gradient
forms comes from the kinetic energy, we shall
take the example of $H=-\Delta$ in the Hilbert space
of square integrable functions $L^2(\mathbb{R}^3)$.
What is the domain of $H$? In fact, many domains are
possible. Let us be very conservative and take the domain
$D_\rmmin$ to be the set of infinitely differentiable functions
with compact support (i.e. which are zero outside a
bounded domain of $\mathbb{R}^3$).
It is clear that $D_\rmmin$ in included in $L^2(\mathbb{R}^3)$ and that
$H$ is well defined on $D_\rmmin$. However this is not the
domain that we want because the domain of the adjoint of
$H$ is not $D_\rmmin$.

The Hamiltonian $H$ with domain $D_\rmmin$
has the adjoint $H^\dagger$ with domain $D(H^\dagger)$
defined as follows (ref. \onlinecite{ReedSimonI}, p.252).
For any $\psi\in D(H^\dagger)$ and $\phi\in D_\rmmin$,
$\int \phi^*(\bfr) \big(H^\dagger \psi(\bfr)\big)\dd\bfr=
-\int \big(\Delta \phi(\bfr)\big)^* \psi(\bfr)\dd\bfr$. 
The function $\phi$ is smooth and with compact support, so
the previous relation shows that
$H^\dagger \psi(\bfr)=-\Delta  \psi(\bfr)$ in the sense
of distributions. As such, $\psi$ does not need to
be continuous for $\Delta  \psi(\bfr)$ to have a meaning
in the sense of distributions. But $H^\dagger$ is
an operator from $\Hcal$ to $\Hcal$, therefore
$\Delta  \psi(\bfr)$ must belong to $\Hcal$.
To summarize, the adjoint of  $H$ with domain $D_\rmmin$
is the operator $-\Delta$ acting in the sense of 
distributions, and its domain is 
the set of square integrable functions $\psi$
such that $\Delta\psi$ is square integrable.
It can be shown that the domain of $H^\dagger$
is the second Sobolev space \cite{ftn3}
$W_2$  (ref. \onlinecite{ReedSimonII}, p.55).

From this example, it is clear that the domain of $H$ and
of $H^\dagger$ can be different. An operator is called self-adjoint when
$H=H^\dagger$ and the domain of $H$ is equal to the domain of
$H^\dagger$. We want self-adjoint operators
because many properties that we take for granted in quantum physics
(Rayleigh-Ritz principle, expansion over eigenstates, 
unitarity of $\exp(itH)$, etc.) are valid only for self-adjoint
operators (ref. \onlinecite{ReedSimonI}, p.256).

Is there a domain over which $H=-\Delta$ is self-adjoint?
Yes, and this domain is unique, it is $W_2$ 
(ref. \onlinecite{ReedSimonII}, p.54).
This unicity is not general. For example, the radial Schr\"odinger operator
\begin{eqnarray}
h &=& -\frac{d^2}{dr^2} -\frac{2}{r}\frac{d}{dr}
+ \frac{\ell(\ell+1)}{r^2},
\end{eqnarray}
has only one self-adjoint domain when 
$\ell\not=0$, but has a family of self-adjoint
domains when $\ell=0$ \cite{Bulla}
(technically, this is because the Laplacian is not
essentially self-adjoint in a space of dimension
$n<4$, see \cite{ReedSimonII} p.\~129).

\subsection{Rayleigh-Ritz principle}
The importance of the self-adjoint domain $D(H)$  of $H$ for
the APW method comes from the fact that, if $H$ is
self-adjoint and bounded from below (i.e. has a ground state),
then for any $n$-dimensional subspace $V$ of $D(H)$
with orthonormal basis \cite{ftn4}
$\phi_i$, the $k$-th eigenvalue
$\lambda_k$ of the matrix $\langle \phi_i|H|\phi_j\rangle$ is larger
than the  $k$-th eigenvalue $E_k$ of $H$. Moreover, 
for realistic Hamiltonians,
it can be shown that $\lambda_0$ converges to the ground state energy
$E_0$ of $H$ as the dimension $n$ of $V$ increases 
(ref. \onlinecite{ReedSimonIV} p.82). In favorable cases
all the eigenvalues $\lambda_k$ converge to  $E_k$.
This happens in particular when the eigenvalues are discrete
\cite{Fix,Volkmer}, which is
the case of the Bloch energies $E_n(\bfk)$ for a given 
Brillouin zone vector $\bfk$.

In other words, if the basis functions $\phi_i$ belong
to $D(H)$, the Rayleigh-Ritz principle gives an approximate
ground state with an energy greater than the true ground state,
and the approximate ground state converges to the true ground state
by increasing the number of basis functions. This is exactly
the properties we assume when using the Rayleigh-Ritz principle.
It seems that the question of the boundary conditions is solved
by determining which basis functions belong to $D(H)$.
However, if the derivative of a basis function $\phi$ is not continuous
at the sphere boundary, then the second derivative involved
in the condition $\Delta\phi\in L^2(\mathbb{R}^3)$ brings a
Dirac $\delta$ function at the sphere radius, and the square
of the $\delta$ function is not defined, so  $\Delta\phi$
does not belong
to $L^2(\mathbb{R}^3)$. Therefore, for a basis function to
be in $D(H)$, it must be continuous with continuous derivatives
at the sphere boundary.

\section{Relaxing the boundary condition}
The previous result is disappointing because the condition of
a continuous derivative at the boundary seems much too strong. 
Many successful calculations of solid-state
properties were carried out with basis functions that are only 
continuous at the sphere boundary. There should be a way
to enlarge the domain of the basis function.
This way exists, it is called the theory of quadratic forms
(ref. \onlinecite{ReedSimonI} p.276).

\subsection{Quadratic forms}
We consider only the quadratic form associated with a self-adjoint
operator $H$, and more precisely with $H=-\Delta$.
This quadratic form is defined by
\begin{eqnarray}
Q(\phi,\psi) &=& \langle \phi| H \psi\rangle
=-\int \phi^*(\bfr) \Delta \psi(\bfr),
\label{Q1}
\end{eqnarray}
for any $\phi$ and $\psi$ in $D(H)$.
In general, $Q$ can be extended to a domain
$Q(H)$ larger than $D(H)$. To increase the domain of
$Q$ we rewrite the quadratic form as
\begin{eqnarray}
Q(\phi,\psi) &=& \int \nabla\phi^*(\bfr) \cdot \nabla \psi(\bfr).
\label{Q2}
\end{eqnarray}
The transformation from (\ref{Q1}) to (\ref{Q2}) is valid
if $\phi$ and $\psi$ are in $D(H)$.
Now the form (\ref{Q2}) has a larger domain 
$Q(H)$ defined as the set of 
$\phi(\bfr)\in L^2(\mathbb{R}^3)$ such that
$\nabla\phi(\bfr)\in L^2(\mathbb{R}^3)$.
This is a Hilbert space called the first Sobolev space $W_1$
(ref. \onlinecite{ReedSimonII} p.50).
This way of enlarging the domain of $Q$ is mathematically unambiguous:
equation (\ref{Q1}) defines a closable quadratic form on $D(H)$ and its
(unique) closure is the quadratic form $Q$ given by equation (\ref{Q2})
with domain $Q(H)=W_1$ (ref. \onlinecite{ReedSimonII} p.177).

\subsection{Rayleigh-Ritz principle for quadratic forms}
This extension of the domain of $H$ is interesting only
if we can use the Rayleigh-Ritz principle. 
It is not difficult to show that this is the case
\cite{ftn5}:
For any $n$-dimensional subspace $V$ of $Q(H)$
with orthonormal basis $\phi_i$, the $k$-th eigenvalue
$\lambda_k$ of the matrix $\langle \phi_i|H|\phi_j\rangle$ is larger
than the  $k$-th eigenvalue $E_k$ of $H$. 
In favorable cases, $\lambda_k$ converge to $E_k$ by
increasing the dimension of $V$ \cite{Volkmer}.

We can now determine the regularity of the basis functions.
If they are continuous at the sphere boundary, the derivative
along the sphere radius has a step, which is a square integrable 
function. Therefore,
the APW basis functions are in $Q(H)$ if they are continuous.
We can even prove that they have to be continuous. 
This is done with Sobolev's lemma
(ref. \onlinecite{ReedSimonII} p.52).
However, we must not consider the Sobolev space
$W_1(\mathbb{R}^3)$ but the  Sobolev space
$W_1(\mathbb{R})$ because the use of radial wave functions
transforms the three-dimensional gradient into a one-dimensional
gradient. Inside sphere $S_\alpha$, the radial function
is $f_{i\ell m}(r_\alpha)$ (see note \onlinecite{ftn1}),
outside this sphere an expansion of the planewave over spherical
harmonics shows that the radial function is proportional to
the Bessel function $j_\ell(k r_\alpha)$.
We want the gradient of this radial function to be
square integrable, so the radial function is in $W_1(\mathbb{R})$.
Sobolev's lemma tells us
now that all functions of $W_1(\mathbb{R})$ are continuous,
so the radial functions are continuous in $Q(H)$.

\subsection{Realistic Hamiltonians}
In the foregoing sections, we dealt only with the operator
$-\Delta$, whereas interesting Hamiltonians have
the form $H=-\Delta+V$, where $V$ is a real potential, which
is smooth except for Coulomb singularities at the nuclei.
In an atom or a molecule, $V$ goes to zero at infinity,
in a crystal $V$ is periodic. It turns out that the addition
of a potential does not modify our conclusions.

For an atom or a molecule, the potential $V$ 
is the sum of a finite number of Coulomb potentials
plus a bounded and square integrable potential
(e.g. the exchange and correlation potentials in the
local density approximation to the density functional theory).
For these potentials, $H$ is self-adjoint on the same domain
as $-\Delta$ (ref. \onlinecite{ReedSimonII} p. 165).
Thus, the Coulomb singularities do not change the domain
\cite{ftn6}. 
Moreover, the form domain is the same as that of $-\Delta$
(this is essentially because the difference between
$D(H)$ and $Q(H)$ comes from the fact that the first
one uses $\Delta\phi$ whereas the second one uses $\nabla\phi$,
and this does not depend on $V$ if the singularities
of $V$ are not stronger than the Coulomb potential \cite{ftn7}). 
Therefore, the whole discussion is still valid for realistic
Hamiltonians representing an atom or a molecule.

For a solid, similar results are available. If $\Omega$
is a primitive cell of the crystal, the Hilbert space
$\Hcal_\bfk$ for the Brillouin zone vector $\bfk$ is the
set of functions $f(\bfr)$ that are square integrable on $\Omega$
and satisfy the Bloch conditions 
$f(\bfr+\bfR)=\exp(i\bfk\cdot\bfR) f(\bfr)$
and $\nabla f(\bfr+\bfR)=\exp(i\bfk\cdot\bfR) \nabla f(\bfr)$,
for any point $\bfr$ on the surface of $\Omega$
and any lattice vector $\bfR$ such that $\bfr+\bfR$
is also on the surface of $\Omega$
(ref. \onlinecite{ReedSimonIV} p. 303).
If $H(\bfk)$ is the Hamiltonian $-\Delta+V$
acting on $\Hcal_\bfk$ and if the periodic potential $V$ is
smooth except for a finite number of Coulombic singularities,
the domain of $H(\bfk)$ is the Sobolev space
$W_2(\Omega)$, i.e. the set of functions $\phi$ of
$\Hcal_\bfk$ such that $\Delta\phi$ is square integrable
on $\Omega$.  The Bloch conditions ensure that
the quadratic form associated with $H(\bfk)$ is
\begin{eqnarray*}
Q(\phi,\psi) &=& 
-\int_\Omega \phi^*(\bfr) \Delta \psi(\bfr)
+\int_\Omega \phi^*(\bfr) V(\bfr) \psi(\bfr)
\\&=&
\int_\Omega \nabla\phi^*(\bfr) \cdot \nabla \psi(\bfr)
+\int_\Omega \phi^*(\bfr) V(\bfr) \psi(\bfr).
\end{eqnarray*}
for any $\phi$ and $\psi$ in $D(H(\bfk))$.
Again the second expression for $Q$ can be extended
to the form domain $Q(H(\bfk))=W_1(\Omega)$.
Therefore, the whole discussion remains valid in the
presence of a periodic potential and the basis functions
must be continuous at the sphere boundary.

\subsection{Discontinuous functions}
We can try to further relax the boundary conditions and
to take basis functions which are not continuous at the
sphere boundary. This was proposed by a number of authors
\cite{Leigh,Hirschfelder,Snyder,Schlosser,Loucks,Weare,%
Hall67,Hall69,Hall70,McCavert,Silverstone,Antoci,Ludena75,%
Stuebing,Inglesfield,Brownstein,Trail,Brownstein02,Szmytkowski}

Among practitioners of APW methods, the prevailing opinion is that
discontinuous boundary conditions are legitimate. For example
Shaughnessy and coll. \cite{Shaughnessy} state that 
``There is no {\it{a priori}} 
reason why the different forms of the basis functions, inside 
and outside the spheres, should match in either value or slope 
at the sphere surface''.

We saw that there is a good reason for using continuous
basis functions. This ensures the validity of the Rayleigh-Ritz
principle, from which we can deduce that to each state of the
eigenvalue problem on the reduced vector space $V$ corresponds
a state of $H$, that the eigenvalues of the reduced problem
are larger than the corresponding eigenvalue of $H$
and that convergence to the eigenvalues of $H$ can be achieved
by increasing the dimension of $V$.

One could be optimistic and hope that the Rayleigh-Ritz
principle is still valid for discontinuous functions.
However, such a hope is generally not fulfilled.
This was shown by Brownstein \cite{Brownstein} who investigated 
simple Hamiltonians $H$ with discontinuous 
basis functions. He showed that discontinuity can lead to
rather unpleasant consequences: (i) spurious states appear,
for example states with energies smaller than the ground
state of $H$, (ii) increasing the dimension of the
reduced vector space $V$ does not always improve the agreement
with the true eigenvalues, (iii) the degeneracy of the
states can be wrong. Moreover, when complex wavefunctions
are used (this is necessary in band-structure calculations),
some formulations give an energy which is a complex 
number \cite{Brownstein02}. These phenomena are obviously
quite dangerous when one tries to calculate (or predict)
properties of matter and it seems reasonable that 
discontinuous basis functions are not used in practical 
implementations of the APW method. 
Of course, as shown by Brownstein, calculations using
discontinuous basis functions \emph{can} yield correct results
in favorable cases. But no general principle
ensures that the results of the computation are the correct 
energies and eigenstates of $H$.

\section{Conclusion}
The results of this paper are clear-cut: 
for the Rayleigh-Ritz principle to be valid the basis functions
have to be continuous at the sphere boundary and the
gradient form of the kinetic energy operator has to
be used.

However, the practical implementation of this condition
looks suspect because the
basis functions are expanded over spherical harmonics
inside the spheres, up to a maximum angular momentum
$\ell_\rmmax$. No finite expansion over spherical harmonics
can match continuously plane waves outside the spheres.
Therefore, the basis functions seem to be {\emph{discontinuous}}.
This is precisely what motivated Leigh \cite{Leigh} to
investigate the APW method with discontinuous basis functions.

In practice, the APW computer codes do not calculate the additional 
terms derived by Leigh to take the discontinuity into account, 
they use the equations for continuous
basis functions and they truncate the basis.
In mathematical terms,
the basis functions are continuous but the Hamiltonian quadratic 
form $Q$ is replaced by $Q_\ell$ which strips off from $Q$ all spherical
harmonics larger than $\ell$ inside the atomic spheres.
In most applications we are only interested in a
fixed number of lowest eigenvalues, and it is observed 
that the eigenvalues converge with $\ell$. The standard physical
argument for this is the fact that the kinetic energy term
$\ell(\ell+1)/r^2$ becomes very large and hides the influence
of the potential or of the energy \cite{Leigh}. However, as
far as I know, further mathematical
work is required to rigorously confirm this. 

Finally, we would like to come back to Slater's intuition.
We saw that the form domain is the correct set in which
basis functions can be chosen. So we might reformulate
Slater's statement as ``quadratic forms
are more fundamental than self-adjoint operators''.
This is confirmed by another property. We saw that 
a Hamiltonian $H$ can have several domains over which it
is self-adjoint. Only one of these domains is included
in the form domain of the quadratic form corresponding
to $H$ (ref. \onlinecite{ReedSimonII} p. 177). 
Different domains correspond to different boundary conditions
and therefore to different predictions of physical properties.
It turns out that experimental results are given by
the domain selected by the quadratic form \cite{Bulla}. So it seems
that, after all, Slater was right and that quadratic forms are
indeed more fundamental.

\begin{acknowledgments}
I thank Matteo Calandra for fruitful discussions
and a thorough reading of the manuscript.
\end{acknowledgments}


\begin{thebibliography}{10}

\bibitem{Singh}
D.J. Singh.
\newblock {\em Planewaves, Pseudopotentials and the LAPW method}.
\newblock Kluwer Academic Publishers, Boston, 1994.

\bibitem{Slater37}
J.C. Slater.
\newblock Wave functions in a periodic potential.
\newblock {\em Phys. Rev.}, 51:646--51, 1937.

\bibitem{ftn1}
Inside the sphere $S_\alpha$, the basis functions $\phi_i(\bfr)$ are written as
  a sum over $\ell$ and $m$ of $f_{i\ell m}(r_\alpha)
  Y_\ell^m(\theta_\alpha,\phi_\alpha)$, where
  $(r_\alpha,\theta_\alpha,\phi_\alpha)$ are the spherical coordinates of
  $\bfr-\bfR_\alpha$, with $\bfR_\alpha$ the center of $S_\alpha$, and where
  $f_{i\ell m}(r_\alpha)$ is a smooth function of $r_\alpha$ up to the sphere
  radius.

\bibitem{Rayleigh}
J.W. Rayleigh.
\newblock In finding the correction for the open end of an organ-pipe.
\newblock {\em Phil. Trans.}, 161:77, 1870.

\bibitem{Ritz}
W.~Ritz.
\newblock {\"U}ber eine neue {M}ethode zur {L}{\"o}sung gewisser
  {V}ariationsprobleme der mathematischen {P}hysik.
\newblock {\em J. reine angew. Math.}, 135:1--61, 1908.

\bibitem{ftn2}
This amounts to solving the generalized eigenvalue problem $ha=Esa$, where
  $h_{ij}=\langle \phi_i | H | \phi_j\rangle$, $s_{ij}=\langle \phi_i |
  \phi_j\rangle$ and $a_i$ are the coefficients of $\psi=\sum_i a_i \phi_i$.

\bibitem{Andersen}
O.K. Andersen.
\newblock Linear methods in band theory.
\newblock {\em Phys. Rev. B}, 12:3060--83, 1975.

\bibitem{Sjostedt}
E.~Sj{\"o}stedt, L.~Nordstr{\"o}m, and D.J. Singh.
\newblock An alternative way of linearizing the augmented plane-wave method.
\newblock {\em Solid State Commun.}, 114:15--20, 2000.

\bibitem{Madsen}
G.K.H. Madsen, P.~Blaha, K.~Schwarz, E.~Sj{\"o}stedt, and L.~Nordstr{\"o}m.
\newblock Efficient linearization of the augmented plane-wave method.
\newblock {\em Phys. Rev. B}, 64:195134, 2001.

\bibitem{Brownstein}
K.R. Brownstein.
\newblock Bound state variational principle employing a discontinuous trial
  function.
\newblock {\em J. Math. Phys.}, 36:76--85, 1995.

\bibitem{Szmytkowski}
R.~Szmytkowski and S.~Bielski.
\newblock Variational principles for bound states of {S}chr{\"o}dinger and
  {D}irac equations allowing the use of discontinuous trial functions.
\newblock {\em Int. J. Quant. Chem.}, 97:966--76, 2004.

\bibitem{ReedSimonI}
M.~Reed and B.~Simon.
\newblock {\em Methods of Modern Mathematical Physics. I {F}unctional
  Analysis}.
\newblock Academic Press, New York, second edition, 1980.

\bibitem{ftn3}
A function $f$ belongs to the Sobolev space $W_2$ if and only if $f$, $\partial
  f/\partial x^i$ and $\partial^2 f/\partial x^i \partial x^j$ are in
  $L^2(\mathbb{R}^3)$ for all $i$ and $j$ in $\{1,2,3\}$. Although Sobolev
  spaces are Hilbert spaces, this is not contradictory with the
  Hellinger-Toeplitz theorem because $H^\dagger$ is not defined from $W_2$ to
  $W_2$ (i.e. $H^\dagger W_2$ is not in $W_2$).

\bibitem{ReedSimonII}
M.~Reed and B.~Simon.
\newblock {\em Methods of Modern Mathematical Physics. II {F}ourier Analysis,
  Self-adjointness}.
\newblock Academic Press, New York, 1975.

\bibitem{Bulla}
W.~Bulla and F.~Gesztesy.
\newblock Deficiency indices and singular boundary condition in quantum
  mechanics.
\newblock {\em J. Math. Phys.}, 26:2520--8, 1985.

\bibitem{ftn4}
The condition that the functions $\phi_i$ are orthonormal is not essential
  because $V$ is finite dimensional, so it is always possible to orthonormalize
  the basis of $V$ used in practice.

\bibitem{ReedSimonIV}
M.~Reed and B.~Simon.
\newblock {\em Methods of Modern Mathematical Physics. IV Analysis of
  Operators}.
\newblock Academic Press, New York, 1978.

\bibitem{Fix}
G.~Fix.
\newblock Orders of convergence of the {R}ayleigh-{R}itz and the
  {W}einstein-{B}azley methods.
\newblock {\em Proc. Nat. Acad. Sci. USA}, 61:1219--23, 1968.

\bibitem{Volkmer}
H.~Volkmer.
\newblock Error estimates for {R}ayleigh-{R}itz approximations of eigenvalues
  and eigenfunctions of the {M}athieu and spheroidal wave equations.
\newblock {\em Constr. Approx.}, 20:39--54, 2004.

\bibitem{ftn5}
Reed and Simon demonstrate the validity of the Rayleigh-Ritz principle for
  basis functions in $D(H)$ by using the min-max principle on $D(H)$ (see ref.
  \onlinecite{ReedSimonIV} p.82). But the min-max principle can be extended to
  $Q(H)$, (see ref. \onlinecite{ReedSimonIV} p.78). It is then straightforward
  to repeat the proof given by Reed and Simon to show that the Rayleigh-Ritz
  principle is valid for basis functions in $Q(H)$.

\bibitem{ftn6}
This can be understood by the following qualitative argument. If a function
  $\phi(r)=r^\alpha$ is such that $\Delta \phi(r)$ is square integrable, then
  $\alpha > 1/2$. Now $\phi(r)$ multiplied by a Coulomb potential has a
  singularity in $r^{\alpha-1}$ which is square integrable. Thus the Coulomb
  singularity does not restrict the domain of $H$.

\bibitem{ftn7}
If a function $\phi(r)=r^\alpha$ is such that $\nabla \phi(r)$ is square
  integrable, then $\alpha > -1/2$. Then $\phi(r)$ multiplied by a Coulomb
  potential has a singularity in $r^{\alpha-1}$ which is square integrable.
  Thus the Coulomb singularity does not restrict the form domain of $H$.

\bibitem{Leigh}
R.S. Leigh.
\newblock The augmented plane wave and related methods for crystal eigenvalue
  problems.
\newblock {\em Proc. Phys. Soc. London A}, 69:388--400, 1956.

\bibitem{Hirschfelder}
J.O. Hirschfelder and G.V. Nazaroff.
\newblock Applicability of approximate quantum-mechanical wave functions having
  discontinuities in their first derivatives.
\newblock {\em J. Chem. Phys.}, 34:1666--70, 1961.

\bibitem{Snyder}
L.C. Snyder and R.G. Parr.
\newblock Problems in perturbation theory calculation of diamagnetic
  susceptibility and nuclear magnetic shielding in molecules. illustration with
  the hydrogen atom.
\newblock {\em J. Chem. Phys.}, 34:837--42, 1961.

\bibitem{Schlosser}
H.~Schlosser and P.M. Marcus.
\newblock Composite wave variational method for solution of the energy-band
  problem.
\newblock {\em Phys. Rev.}, 131:2529--46, 1963.

\bibitem{Loucks}
T.L. Loucks.
\newblock Relativistic electronic structure in crystals. {I}. {T}heory.
\newblock {\em Phys. Rev.}, 139:A1333--7, 1965.

\bibitem{Weare}
J.H. Weare and R.G. Parr.
\newblock A stationary principle for discontinuous trial functions and a
  framework for zero-differential-overlap theories of electronic structure.
\newblock {\em Chem. Phys. Lett.}, 1:349--50, 1967.

\bibitem{Hall67}
G.G. Hall.
\newblock A variation principle for discontinuous wavefunctions.
\newblock {\em Chem. Phys. Lett.}, 1:495--6, 1967.

\bibitem{Hall69}
G.G. Hall, J.~Hyslop, and D.~Rees.
\newblock A minimum principle for atomic systems allowing the use of
  discontinuous wave functions.
\newblock {\em Int. J. Quant. Chem.}, 3:195--204, 1969.

\bibitem{Hall70}
G.G. Hall, J.~Hyslop, and D.~Rees.
\newblock A minimum principle for molecular systems allowing the use of
  discontinuous wave functions.
\newblock {\em Int. J. Quant. Chem.}, 4:5--20, 1970.

\bibitem{McCavert}
P.~McCavert and M.R.H. Rudge.
\newblock On the use of regional wavefunctions in bound-state calculations.
\newblock {\em Proc. R. Soc. London A}, 328:429--444, 1972.

\bibitem{Silverstone}
H.J. Silverstone and E.W. Stueging.
\newblock Kinetic-energy expectation values with discontinuous approximate wave
  functions.
\newblock {\em Phys. Rev. A}, 5:1092--3, 1972.

\bibitem{Antoci}
S.~Antoci.
\newblock The use of discontinuous trial functions in the computation of the
  electronic structure of molecules: {C}alculation of the {ESCA} spectrum of
  tetrathiofulvalene ({TTF}).
\newblock {\em J. Chem. Phys.}, 63:697--701, 1975.

\bibitem{Ludena75}
E.V. Lude{\~n}a.
\newblock Variational principles for discontinuous wave functions and the
  independent particle model of electronic structure.
\newblock {\em Int. J. Quant. Chem.}, 9:1069--85, 1975.

\bibitem{Stuebing}
E.W. Stuebing, J.H. Weare, and R.G. Parr.
\newblock Discontinuous approximate molecular electronic wave-functions.
\newblock {\em Int. J. Quant. Chem.}, 11:81--102, 1977.

\bibitem{Inglesfield}
J.E. Inglesfield.
\newblock A method of embedding.
\newblock {\em J. Phys. C: Solid State Phys.}, 14:3795--806, 1981.

\bibitem{Trail}
J.R. Trail and D.M. Bird.
\newblock Core reconstruction in pseudopotential calculations.
\newblock {\em Phys. Rev. B}, 60:7863--74, 1999.

\bibitem{Brownstein02}
K.R. Brownstein and M.~Zhou.
\newblock Comment on ``{C}ore reconstruction in pseudopotential calculations".
\newblock {\em Phys. Rev. B}, 65:197101, 2002.

\bibitem{Shaughnessy}
D.J. Shaughnessy, G.R. Evans, and M.I. Darby.
\newblock An improved {LAPW} method for the calculation of self-consistent band
  structures.
\newblock {\em J. Phys. F: Met. Phys.}, 17:1671--9, 1987.

\end{thebibliography}
\end{document}